\documentclass[fleqn,usenatbib]{mnras}

\usepackage{newtxtext,newtxmath}
\usepackage[T1]{fontenc}

\DeclareRobustCommand{\VAN}[3]{#2}
\let\VANthebibliography\thebibliography
\def\thebibliography{\DeclareRobustCommand{\VAN}[3]{##3}\VANthebibliography}

\usepackage{graphicx}	% Including figure files
\usepackage{amsmath}	% Advanced maths commands

\title[A disc with magnetic outflows in 1ES~1927+654]{An accretion disc with magnetic outflows triggered by a sudden mass accretion event in changing-look active galactic nucleus 1ES~1927+654}

\author[X. Cao et al.]{
Xinwu Cao,$^{1}$\thanks{E-mail: xwcao@zju.edu.cn (XC)}
Bei You,$^{2}$\thanks{E-mail: youbei@whu.edu.cn (BY)}
and Xing Wei$^{3}$
\\
$^{1}$Institute for Astronomy, School of Physics, Zhejiang University, 866 Yuhangtang Road, Hangzhou, 310058, People's Republic of China\\
$^{2}$Department of Astronomy, School of Physics and Technology, Wuhan University, Wuhan, 430072, People's  Republic of China \\
$^{3}$Department of Astronomy, Beijing Normal University, Beijing, 100875, People's Republic of China}

\date{Accepted 2023 September 18. Received 2023 September 6; in original form 2023 August 1}

\pubyear{2015}

\begin{document}
\label{firstpage}
\pagerange{\pageref{firstpage}--\pageref{lastpage}}
\maketitle

% Abstract of the paper
\begin{abstract}
1ES 1927+654 was known as a type 2 Seyfert galaxy, which exhibited drastic variability recently in ultraviolet (UV)/optical and X-ray bands. An UV/optical outburst was observed in the end of 2017, and it reached the peak luminosity $\sim 50$ days later. The high-cadence observations showed a rapid X-ray flux decline with complete disappearance of the power-law hard X-ray component when the soft X-ray thermal emission reached its lowest level about $150$ days after the UV/optical peak. The power law X-ray component reappeared with thermal X-ray emission brightening from its lowest flux within next $\sim$ 100~days. We assume an episodic accretion event taking place in the outer region of the disc surrounding a central black hole (BH), which is probably due to a red giant star tidally disrupted by the BH. The inner thin disc with corona is completely swept by the accretion event when the gas reaches the innermost circular stable orbit. The field threading the disrupted star is dragged inwards by  the disc formed after the tidal disruption event, which accelerates outflows from the disc. The disc dimmed since a large fraction of the energy released in the disc is tapped into the outflows. The accretion rate of the episodic accretion event declines, and ultimately it turns out to be a thin disc, which is inefficient for field advection, and the outflows are switched off. A thin disc with corona reappears later after the outburst.
\end{abstract}

% Select between one and six entries from the list of approved keywords.
% Don't make up new ones.

%\keywords{}

%%%%%%%%%%%%%%%%%%%%%%%%%%%%%%%%%%%%%%%%%%%%%%%%%%

%%%%%%%%%%%%%%%%% BODY OF PAPER %%%%%%%%%%%%%%%%%%

\section{Introduction} \label{sec:intro}

Active galactic nuclei (AGN) can be classified as type 1 or type 2 according to their emission line properties (width of lines) \citep[][]{1989agna.book.....O}, which is supposed to be caused by the viewing angle of the observer with respect to the axis of the disc surrounding the massive black hole (BH).  This is the essence of the so-called unification scheme  \citep[][]{1993ARA&A..31..473A,1995PASP..107..803U}. However, a small fraction of AGNs have drastic changes in multi-wavebands within a short period of time. Some of them appear as type 1 AGNs when brighten, while they become type 2 in the dim state, namely, changing look (CL) AGNs \citep[see][for a review and the references therein]{2022arXiv221105132R}. These features defy the unification scheme of AGN.

AGN are powered by accretion of gas onto supermassive BHs, and rich features observed in multi-wavebands on CL AGN provide important clues on the physics of the BH accretion, and even the relation with their host galaxies \citep[e.g.,][]{2015ApJ...800..144L,2016A&A...593L...9H,2019ApJ...883L..44G,2020ApJ...890L..29A,2020A&A...643L...7K,2021ApJ...912...20J,2021MNRAS.506.4188L,2021MNRAS.508..144G,2022ApJ...930...46L,2022MNRAS.514..403T,2022ApJ...937...31G,2023ApJ...943...63N,2023MNRAS.519.3903L}. Great efforts have been devoted to search for CL AGNs, and there are more than 100 CL AGNs have been discovered so far \citep[e.g.,][]{1976ApJ...210L.117T,1986ApJ...311..135C,2017ApJ...846L...7S,2018ApJ...862..109Y,2018ApJ...864...27S,2019ApJ...883L..44G,2019Natur.573..381M,2019ApJ...883...94T,2019ApJ...887...15W,2020AJ....159..245W,2022ApJ...933..180G}. A variety of scenarios have been suggested to explain the drastic variability of CL AGNs, among which the most likely explanations are that the CL AGNs are fed by the disc with dramatic changing accretion rate. However, the timescales of the variability in CL AGNs are orders of years or decades, which are much shorter than the timescale of the global accretion rate changes expected in an optically thick, geometrically thin accretion disc surrounding a supermassive BH \citep[see][for the detailed discussion]{2018NatAs...2..102L,2018ApJ...864...27S}. The magnetically elevated disc was suggested by \citet{2019MNRAS.483L..17D}, of which the accretion timescale is significantly reduced, and it could help alleviate the viscosity crisis in CL AGNs \citep[][]{2023arXiv230904092W}. \citet{2021ApJ...916...61F} suggested that the timescale of the inflow gas in the disc can be substantially shortened if most angular momentum of the gas in the disc is removed by magnetically driven outflows \citep[][]{2013ApJ...765..149C}.

1ES~1927+654 was a narrow-line Seyfert galaxy, $z=0.019422$, before the UV/optical outburst observed in 2017 \citep[e.g.,][]{2003A&A...397..557B,2007ApJ...660.1072W,2011ApJ...726L..21T}, while it emerges as an unusual CLAGN since 2017 \citep[][]{2019ApJ...883...94T,2020ApJ...898L...1R,2021ApJS..255....7R,2022ApJ...931....5L,2022ApJ...934...35M}. An ultraviolet (UV)/optical outburst was observed in the end of 2017, and its flux reaches the peak $\sim 50$ days later. Then the UV/optical emission decays with a tidal disruption event (TDE) like light-curve \citep[][]{2019ApJ...883...94T}. Prior to the outburst, the X-ray emission consists of a soft thermal component and a power-law hard X-ray component with a relatively large photon index \citep[][]{2003A&A...397..557B,2013MNRAS.433..421G}. The broad emission lines emerge after the outburst, and their fluxes arrive at the peak $\sim$ 150 days after the UV/optical outburst \citep[][]{2019ApJ...883...94T,2022ApJ...933...70L}. Roughly at the same period of time, their X-ray emission drastically declines about two orders of magnitude to its lowest flux within several ten days \citep[][]{2021ApJS..255....7R}. It is interesting to find that its power-law hard X-ray emission completely disappears when its soft thermal X-ray emission is at the lowest flux, which strongly implies the destruction of the corona \citep[][]{2020ApJ...898L...1R}. The soft thermal X-ray emission increases to the flux prior to the outburst (or even higher) within the next several ten days with the reappearance of the power-law hard X-ray component \citep[e.g.,][]{2021ApJS..255....7R}. A TDE-like UV/optical light-curve strongly implies that the UV/optical outburst originates in a TDE, however, it is difficult to understand why the X-ray emission declines about two orders of magnitude in flux more than one hundred days after the UV/optical outburst \citep[e.g.,][]{2021ApJS..255....7R}, and the X-ray light-curve does not show any characteristics of a normal TDE \citep[][]{2018MNRAS.474.3593K,2016MNRAS.455.2918H}. \citet{2021MNRAS.502L..50S} argued that a change in the accretion rate and an inversion of magnetic flux in a magnetically arrested disc (MAD) are required to explain the extraordinary feature of the X-ray dip after the UV/optical outburst in 1ES~1927+654, though the detailed physical mechanism triggering such a flux inversion event is still quite unclear.

In this work, we propose that an episodic accretion event in the outer region of the disc, most probably a TDE, sweeps up the inner accretion disc with corona. An accretion disc with magnetically driven outflows is being established when the accretion front of the event moves inwards. As a large fraction of the angular momentum and the gravitational energy of the gas in the disc is carried away by the outflows, the radiation of the disc is substantially reduced, and the corona above the inner disc is destructed by the inflow gas triggered by the TDE. This leads to a dip of the soft X-ray emission. The basic observational features of this source are summarized in Section \ref{sec:obs_features}. We describe the model in Section \ref{sec:model}. The last section contains a discussion of the model.

\section{Observational features}\label{sec:obs_features}

We briefly summarize the observational features to be used for constructing a theoretical model for 1ES~1927+654 in this section. The source was observed in 2011 as a Seyfert 2 galaxy with its X-ray luminosity $L_{\rm X}^{0.3-10\rm keV}=6.81\times 10^{42}~{\rm erg~s}^{-1}$ in the energy band of $0.3-10$~keV \citep[][]{2013MNRAS.433..421G}.
A UV/optical outburst was observed in the end of 2017, and it reached the peak luminosity around 50 days later \citep[][]{2019ApJ...883...94T}. The UV/optical peak emission is about four magnitudes brighter than that in the pre-outburst state. During this period of time, no X-ray observation has been performed \citep[][]{2021ApJS..255....7R,2023MNRAS.521.3517H}. The UV/optical emission declined slowly with time from the peak, which is similar to a typical light-curve of a TDE \citep[][]{2019ApJ...883...94T,2022ApJ...931....5L}. The high-cadence observations in X-ray bands have been carried out about several ten days after the UV/optical peak, which showed a rapid X-ray flux decline with the power-law component completely disappearing while the X-ray emission reached its lowest level (about two orders of magnitude lower than that before the X-ray dip) \citep[][]{2021ApJS..255....7R}. With the soft X-ray emission increasing from the lowest flux, the power law X-ray emission reappeared and backed to (or even brighter than) the flux before the X-ray dip within the next $\sim$100~days \citep[][]{2021ApJS..255....7R,2022ApJ...931....5L}.

\section{Model}\label{sec:model}

The UV/optical emission declines with time in the form similar to a typical TDE light-curve after the peak of the outburst, which strongly implies that the UV/optical outburst is triggered by a TDE \citep[e.g.,][]{2019ApJ...883...94T,2020ApJ...898L...1R,2021ApJS..255....7R,2022ApJ...931....5L,2022ApJ...934...35M}. For a normal solar-like star captured by a massive BH, its tidal disruption radius is very close to the innermost circular stable orbit (ISCO). For 1ES 1927+654, a Seyfert 2 galaxy in the pre-outburst period, its inner accretion disc with corona would be destructed instantly when a normal star is tidally disrupted by the BH. The radiation from the inner region of a newly formed TDE disc surrounding a BH with $\sim 10^6~M_\odot$ should dominate in the soft X-ray energy band, whereas no similar slowly declining light-curve in X-ray energy bands has been observed as those observed in normal TDEs \citep[e.g.,][]{2016MNRAS.455.2918H,2018MNRAS.474.3593K}.

\begin{figure}
	\centering
	\includegraphics[width=1.0\columnwidth]{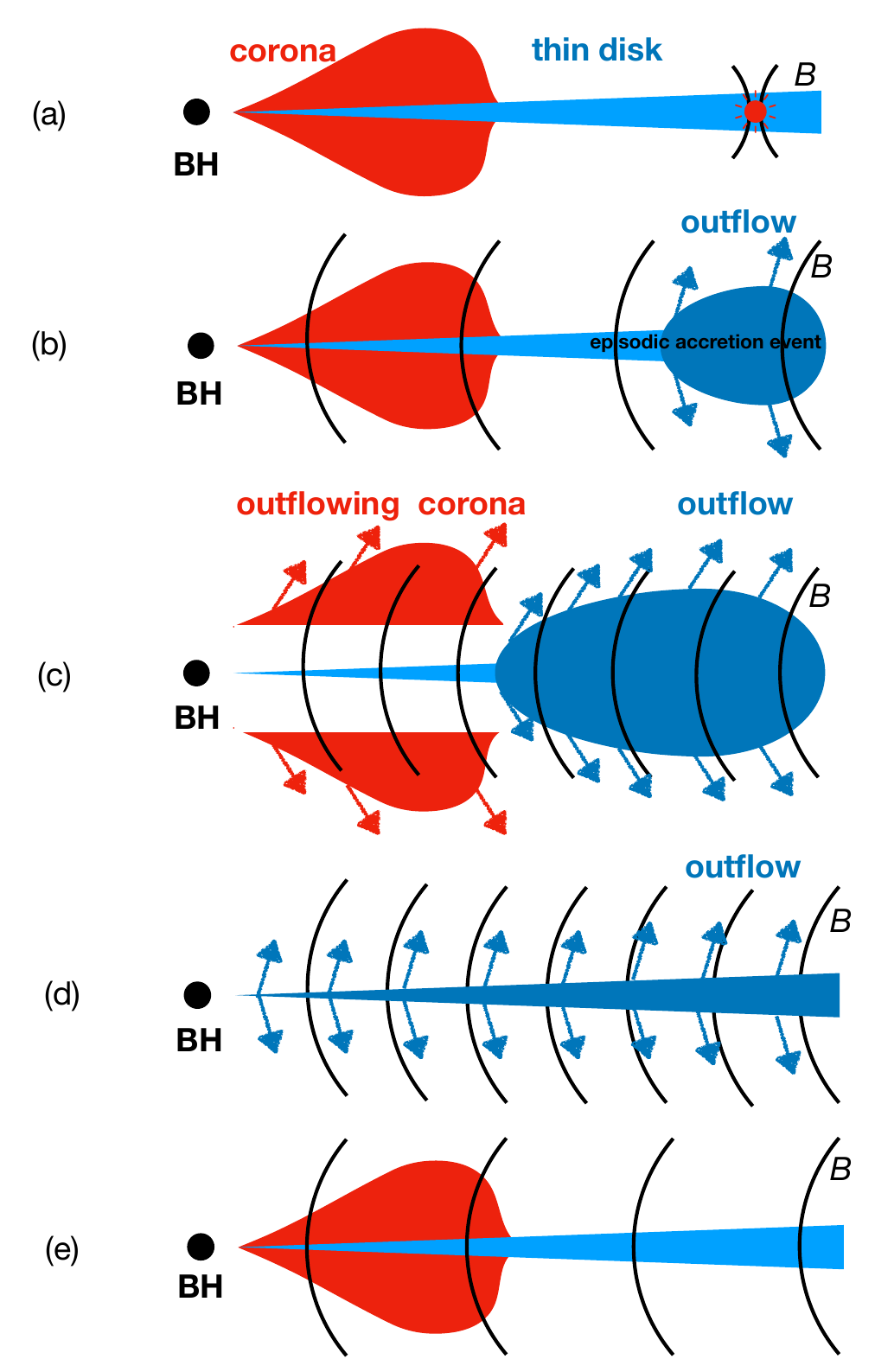}
	\caption{Illustration of the model (not to scale in size), a. a red giant star is being tidally disrupted by the BH; b. a magnetized outer disc is formed after the TDE; c. the hot gas in the corona is being magnetically driven away from the disc; d. the gas front of the episodic accretion event reaches the ISCO, which corresponds to the lowest flux of the X-ray dip observed in July 2018; e. the mass accretion rate declines to a low rate, so the field advection becomes inefficient, and a normal thin disc with corona reappears. }
	\label{fig:model}
\end{figure}

We assume that an episodic accretion event takes place in the outer region of the disc surrounding a central BH, which is triggered by a tidally disrupted star (illustrated in Figure \ref{fig:model}). If this is the case, it requires the disrupted star with a radius much larger than the solar radius or/and much more massive than the Sun (panel a in Figure \ref{fig:model}) \citep[see Equation 1 in][]{1988Natur.333..523R}. An alternative possibility is the BH mass is substantially lower than $10^6M_\odot$ (e.g., the tidal radius $R_{\rm t}\sim 500 R_{\rm g}$ for a solar-like star disrupted by a $2\times 10^4 M_\odot$ BH). It is quite unlikely that 1ES 1927+654 contains such a small BH \citep[see][for the detailed discussion]{2022ApJ...933...70L}. As the relative number of the stars decreases significantly with increasing mass \citep[][]{2001MNRAS.322..231K}, we conjecture it highly possible that a red giant star with solar mass is disrupted in the outer region of the accretion disc in this source.

The UV/optical outburst was observed at the end of 2017 for 1ES 1927+654, and the peak of the UV/optical emission is about four magnitudes brighter than that in the pre-outburst state, which implies that the accretion rate of the newly formed outer disc after the TDE is $\sim 40$ times of that of the primary inner thin disc. As no X-ray observations at the time the UV/optical outburst taking place, one may assume the X-ray emission of the source is roughly similar to the previous observations carried out in 2011, of which the X-ray luminosity is $L_{\rm X}^{0.3-10\rm keV}=6.81\times 10^{42}~{\rm erg~s}^{-1}$ consisting of a soft thermal component and a power-law hard component \citep[][]{2013MNRAS.433..421G}. The accretion rate in the outer disc is sharply increased due to the episodic accretion event, which moves inwards to sweep up the inner accretion disc corona.

The size of a solar mass red giant can be substantially larger than the 
%at its final stage could be as large as $\sim 100-1000$ 
solar radius \citep[][]{1980FCPh....5..287T}. It is known that convection plays an important role in the radial energy transport of a red giant star, which is very similar to the Hayashi track of the proto-stars in the Hertzsprung-Russell diagram. It has a very strong magnetic field  \citep[][]{2007ApJ...671..802B,2015Sci...350..423F}. Thus, the gas in the disc formed after the disruption of a red giant star must be strongly magnetized, and the magnetic field is then dragged inwards by this newly formed disc. The field advection in this disc is efficient due to its large accretion rate/radial velocity, and therefore the outflows will be driven by the field threading the disc (panel b in Figure \ref{fig:model}).

The outer disc with outflows connecting the inner thin disc with corona at a certain radius, which decreases with time as the gas of the episodic accretion event flowing into the inner disc. The strength of the large-scale magnetic field in the inner region of the thin disc also increases with decreasing of the gas front of the episodic accretion event, because the field is maintained by the currents mostly in the outer disc \citep[see][for the detailed explanation]{2021A&A...654A..81C}.  The strength of the field in the thin disc-corona region increases substantially with the accretion front moving to a small radius, which may accelerate the hot gas in the corona away from the disc surface (panel c in Figure \ref{fig:model}). It implies that the emission of the corona may substantially be suppressed before the gas of the episodic accretion event arrives at the ISCO, which is consistent with the fact that the hard X-ray component disappears before the X-ray emission reaches the lowest flux \citep[][]{2021ApJS..255....7R}. The inner thin disc ultimately disappears when the gas of the accretion event approaches the ISCO of the BH. At this moment, a disc accreting at a high rate with magnetically driven outflows is established. The outflows carry away most of the kinetic energy and angular momentum of the gas in the disc \citep[][]{2006A&A...447..813F}, and therefore the radiation of the disc is substantially reduced \citep[][]{2013ApJ...765..149C,2016ApJ...833...30C}. This corresponds to the observed lowest fluxes during the X-ray dip (panel d in Figure \ref{fig:model}). 

As the accretion rate of the episodic accretion event declines with time, the episodic accretion disc evolves to a thin disc when the accretion rate is sufficiently low. The radial velocity of the gas in the thin disc is too low to accumulate the field flux of the gas in the outer region \citep[][]{1994MNRAS.267..235L}. Thus, the field in the inner region of the disc is much weaker than that of the disc when the accretion rate is high in the early stage of the episodic accretion event, which is too weak to accelerate strong outflows. A normal accretion disc with corona is re-established again (panel e in Figure \ref{fig:model}), which can naturally account for the reappearance of the power law X-ray emission about 200 days after the UV/optical outburst.

\subsection{Black hole mass}\label{sec:bh_mass}

The mass of the central BH in 1ES~1927+654 has been estimated from the host galaxy properties, $M_{\rm bh}=1.38_{-0.73}^{+1.37}\times 10^6~M_\odot$, $M_{\rm bh}=3.56_{-0.35}^{+0.38}\times 10^6~M_\odot$, or $M_{\rm bh}=1.08_{-0.81}^{+3.21}\times 10^6~M_\odot$, with different approaches \citep[see][for the details, and the references therein]{2022ApJ...933...70L}. The photon index of the hard X-ray continuum spectrum is found tightly correlated with the Eddington ratio for AGN samples \citep[e.g.,][]{2004ApJ...607L.107W,2008ApJ...682...81S}, which provides an estimate of the BH mass when the hard X-ray photon index and the bolometric luminosity are well measured \citep[][]{2008ApJ...682...81S}. We can estimate the Eddington ratio of the sources using the empirical relation of the Eddington ratio $\lambda$ with the photon index $\Gamma$ \citep[see Eq. 2 in][]{2008ApJ...682...81S},
\begin{equation}
\log \lambda=0.9\pm0.3\Gamma-2.4\pm0.6,
    \label{lambda}
\end{equation}
where $\lambda=L_{\rm bol}/L_{\rm Edd}$, $L_{\rm bol}=\kappa_{\rm x}L_{\rm x}^{0.3-10~{\rm keV}}$, and $\kappa_{\rm x}$ is the bolometric correction factor. The observations in 2011 show that the luminosity in 0.3-10~keV $L_{\rm x}^{0.3-10~{\rm keV}}=6.81\times 10^{42}~{\rm erg~s}^{-1}$ with the hard X-ray photon index in the range of $\Gamma=2.3-2.65$ \citep[][]{2013MNRAS.433..421G,2021ApJS..255....7R}. Thus, the estimated BH mass is in the range of $M_{\rm bh}\sim 0.71-1.48\times 10^5\kappa_{\rm x}~M_\odot$. For a BH with $M_{\rm bh}\sim 10^{5-6}~M_\odot$, the radiation of an optically thick disc dominates in the soft X-ray band, which implies that $\kappa_{\rm x}$ must not deviate much from unity.

The X-ray observations of this source indicate that a blackbody component is always present, though its X-ray continuum spectra show violent variability. Although it looks like the soft X-ray excess ubiquitously observed in quasars \citep[][]{2004MNRAS.349L...7G,2012MNRAS.420.1848D}, however, there are a variety of arguments against the soft X-ray emission observed in this source being the soft excess as observed in quasars \citep[see][for the detailed discussion]{2022ApJ...931....5L}. The typical temperature of the inner region of the disc surrounding a BH with $M_{\rm bh}\sim 10^{5-6}~M_\odot$ ranges from several ten to hundreds eV, which is consistent with the blackbody temperature of the observed soft X-ray emission. In this work, we assume the soft X-ray emission originates in the inner region of the disc in this source. The size of the soft X-ray emission region can be roughly estimated with the temperature and luminosity, which gives only $\sim R_{\rm g}$, if $M_{\rm bh}=10^6~M_\odot$ is adopted. This may imply a smaller BH with $<10^6~M_\odot$ or/and a larger BH spin parameter $a$ \citep[][]{2021ApJS..255....7R}. The observations of the 1~keV reflection features set constraints on the spin parameter, $a\simeq 0.4$ \citep[][]{2021ApJS..255....7R}, or $a\simeq 0.9$ \citep[][]{2022ApJ...934...35M}. Motivated by the aforementioned discussion, we adopt a rather small BH mass $M_{\rm bh}=3\times 10^5~M_\odot$ in our model calculations, which roughly corresponds to the lower limit of the BH mass estimate $M_{\rm bh}=1.08_{-0.81}^{+3.21}\times 10^6~M_\odot$ in \citet{2022ApJ...933...70L}.

\subsection{Impact of an episodic accretion event on the inner accretion disc with corona}\label{sec:impact_acc}

Before the episodic accretion event, a standard thin accretion disc with corona surrounds a massive BH in 1ES~1927+654. The episodic accretion event takes place at the radius $R_{\rm epi}$ with an initial mass accretion rate $\dot{M}_{\rm epi}$, which is much higher than the rate $\dot{M}$ of the inner thin disc. The structure of the inner accretion disc corona remains unchanged at the moment when the episodic accretion event happens. The accretion front of the episodic event moves inwards, and pushes the inner accretion disc, which shrinks in size accordingly. Assuming that the inner disc corona is swept up in a timescale of $\delta\tau$, we estimate the momentum change of the inner disc during the period $\delta\tau$ as
\begin{equation}
 \delta p\sim M_{\rm d}(R<R_{\rm epi})v_{\rm d}, \label{delta_p}
\end{equation}
where $M_{\rm d}(R<R_{\rm epi})$ is the mass of the inner disc encircled by radius $R_{\rm epi}$, and the characteristic shrinking velocity of the inner disc $v_{\rm d}\sim R_{\rm epi}/\delta\tau$. Such momentum change of the inner disc is caused by the inflow gas triggered by the episodic accretion event, so the momentum conservation requires
\begin{equation}
   \dot{M}_{\rm epi}v_{\rm epi}(R_{\rm epi})\delta\tau\sim M_{\rm d}(R<R_{\rm epi})v_{\rm d}\sim M_{\rm d}(R<R_{\rm epi})R_{\rm epi}/\delta\tau, \label{moment_conser}
\end{equation}
which leads to
\begin{equation}
\delta\tau\sim\left[{\frac {M_{\rm d}(R<R_{\rm epi})R_{\rm epi}}{\dot{M}_{\rm epi}v_{\rm epi} }}   \right]^{1/2}.\label{delta_tau}
\end{equation}
The mass of the inner disc is estimated with the standard thin disc model \citep[][]{1973A&A....24..337S}, and we have
\begin{equation}
    {\frac {M_{\rm d}(R<R_{\rm epi})}{\dot{M}_{\rm epi}}}=7.8\times 10^{-14}\alpha^{-1}m\dot{m}^{-1}\dot{m}_{\rm epi}^{-1}r_{\rm epi}^{7/2},~~{\rm day} \label{rat_md_mdot}
\end{equation}
where $\alpha$ is the viscosity parameter. It implies that the inner disc is more heavier for a lower accretion rate, because the radial velocity of the disc $v_{R}\propto \dot{m}^2$, and therefore the surface density of the disc $\Sigma \propto \dot{m}^{-1}$ \citep[][]{1973A&A....24..337S}. The dimensionless quantities are defined as
\begin{displaymath}
m={\frac {M_{\rm bh}}{M_\odot}}, ~~~~
r={\frac {R}{R_{\rm g}}},~~~~~R_{\rm g}={\frac {GM_{\rm bh}}{c^2}},~~~~~\dot{m}={\frac {\dot{M}}{\dot{M}_{\rm Edd}}},
\end{displaymath}
\begin{displaymath}
\dot{M}_{\rm Edd}={\frac {L_{\rm Edd}}{0.1c^2}},~~~{\rm and}~~~
L_{\rm Edd}=1.25\times 10^{38}m~{\rm erg~s}^{-1}.\label{dimensionless}
\end{displaymath}
As the accretion rate of the episodic event $\dot{m}_{\rm epi}>1$, the radial velocity of the inflow gas should be estimated in the frame of a slim disc model. In this work, we adopt the supercritical accretion disc model developed by \citet{2022ApJ...936..141C} to calculate $v_{\rm epi}(r_{\rm epi})$. Then the timescale $\delta\tau$ is estimated with Equation (\ref{delta_tau}), which is corresponding to the time delay of the X-ray dip to the peak of the UV/optical outburst.

The radiation efficiency of a thin disc surrounding a non-rotating BH is $\sim 0.1$, while it can be up to $0.3-0.4$ for an extreme Kerr BH \citep[][]{1974ApJ...191..499P}. If we assume the radiation efficiency $\eta=0.2$, and $\kappa_{\rm X}=1.5$ to account for the emission below $0.3$~keV and above $10$~keV, the dimensionless mass accretion rate $\dot{m}\simeq 0.15$ is estimated from the observed X-ray luminosity $L_{\rm X}^{0.3-10\rm keV}=6.81\times 10^{42}~{\rm erg~s}^{-1}$ for 1ES~1927+654 before the UV/optical outburst. Here we assume the X-ray emission before the outburst is at the same level as that observed in 2011 \citep[][]{2013MNRAS.433..421G}.

It was found that the peak emission of the UV/optical outburst is about four magnitudes brighter than that in the pre-outburst state \citep[][]{2019ApJ...883...94T}, which implies the mass accretion rate of the gas inflow intriggered by the episodic accretion event is $\sim 40$ times of that in the pre-outburst state, i.e., $\dot{m}_{\rm epi}\sim 40\dot{m}\simeq 6$, if the UV/optical photons are emitted from the episodic accretion disc at the outburst peak. Thus, we plot the time delay of the X-ray dip to the UV/optical outburst varying with the location of the episodic accretion event in Figure \ref{fig:fig1} using Equations (\ref{delta_tau}) and (\ref{rat_md_mdot}).
%where  $v_{\rm epi}$ is derived with the disc model calculation in \citet{2022ApJ...936..141C}.  
As $\delta\tau\simeq 150$~days constrained by the observations (see Section \ref{sec:obs_features}), we find that $r_{\rm epi}\simeq 350$, 480, and 650 for $\alpha=0.03$, 0.1, and 0.3, respectively.

\subsection{TDE triggered accretion event}\label{sec:TDE_acc}

\begin{figure}
	\centering
	\includegraphics[width=1.0\columnwidth]{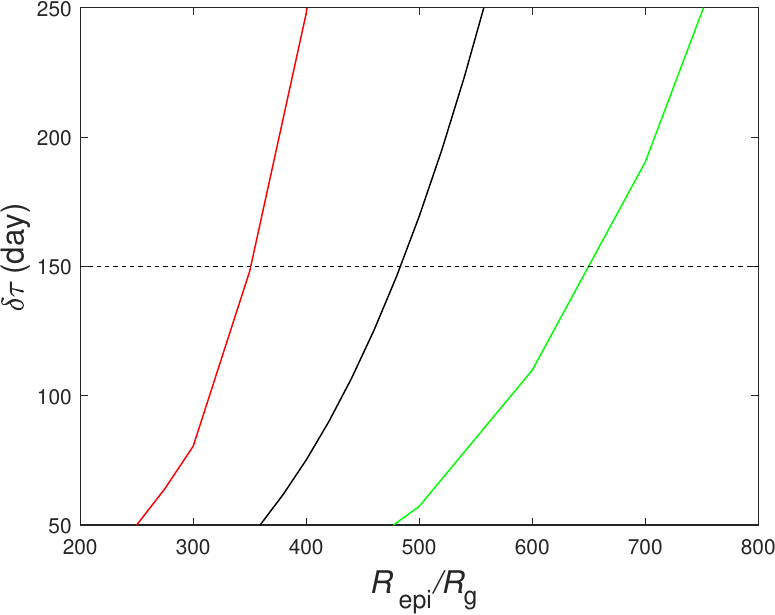}
	\caption{The location of the episodic accretion event versus the time delay of the X-ray emission to the UV/optical outburst. The colors of the lines denote the results with different values of the viscosity parameter, i.e., $\alpha=0.03$ (red), 0.1 (black), and 0.3 (green), respectively.   }
	\label{fig:fig1}
\end{figure}

We conjecture that the UV/optical flare is caused by a sharp increase of the accretion rate in the outer region of the disc, which predominantly radiates UV/optical photons. In Section \ref{sec:impact_acc}, we estimate the location of such an episodic accretion event should be at several hundred gravitational radii as constrained by the observations (see Figure \ref{fig:fig1}).

There is strong evidence that the episodic accretion event is triggered by a TDE in this source \citep[][]{2019ApJ...883...94T,2020ApJ...898L...1R,2022ApJ...931....5L,2022ApJ...934...35M}. The tidal disruption radius of a star captured by a massive BH is
\begin{equation}
r_{\rm t}\equiv {\frac {R_{\rm t}}{R_{\rm g}}}=33.3\left({\frac {M_{\rm bh}}{10^6 M_\odot}}\right)^{-2/3}\left({\frac {R_*}{R_\odot}}\right)\left({\frac {M_*}{M_\odot}}\right)^{-1/3},
    \label{r_t}
\end{equation}
where $M_*$ and $R_*$ are the star mass and radius respectively \citep[][]{1988Natur.333..523R}. For a solar-mass main sequence star, it will be disrupted by a massive BH with $\sim 10^6~M_\odot$ in the region very close to the ISCO, which implies the inner region of the disc will be destructed instantly when the TDE occurs. An X-ray outburst is expected when a new disc is formed after the TDE, however, this has not been observed in 1ES~1927+654. 

\begin{figure}
	\centering
	\includegraphics[width=1.0\columnwidth]{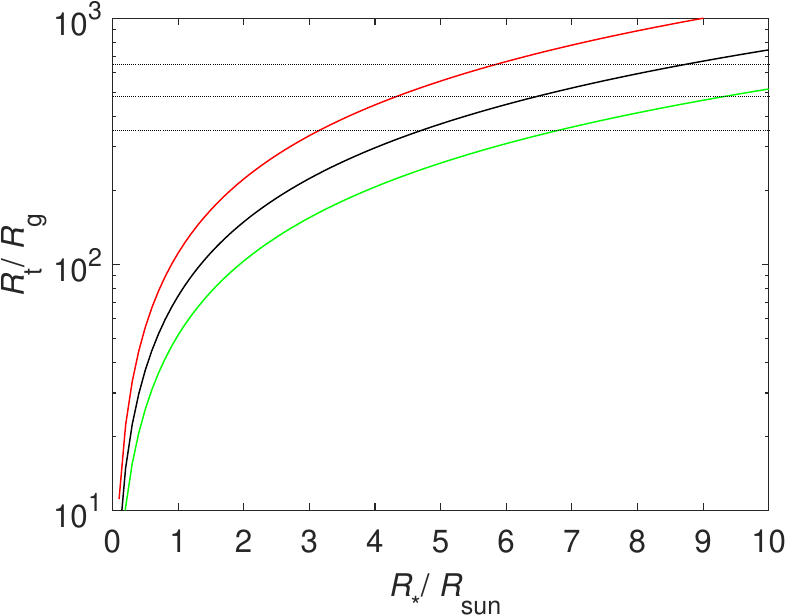}
	\caption{The tidal disruption radius $R_{\rm t}$ versus the star radius with different star masses (red: $0.3M_\odot$, black: $1M_\odot$, and green: $3M_\odot$). The horizontal dotted lines denote $R_{\rm t}=350R_{\rm g}$, $480R_{\rm g}$, and $650R_{\rm g}$, respectively.  }
	\label{fig:fig2}
\end{figure}

We plot the tidal radius of a star varying with the star radius in Figure \ref{fig:fig2} (Equation \ref{r_t}), which implies that the disrupted star must be large in size, if such an episodic accretion event is attributed to a TDE. As the relative number of massive stars decreases rapidly with increasing mass \citep[][]{2001MNRAS.322..231K}, so the probability of a massive star captured by the BH is very low. We suggest a solar-mass red giant star with several times of solar radius is disrupted in the outer region of the accretion disc in this source.

\subsection{Accretion disc with magnetically driven outflows}\label{sec:acc_outflow}

The red giant star evolves in a branch parallel with that of a proto-star in the Hertzsprung-Russell diagram. The interior energy transport in the envelope of a red giant star is convection dominant, which is similar to a proto-star. The convective envelope in a differentially rotating star helps field generation via the dynamo process \citep[][]{1971ApJ...164..491P,1974ApJ...193..419L}, and therefore the field of a giant red star is very strong \citep[][]{2007ApJ...671..802B,2008ApJ...684L..29N}. A strong coherent magnetic field may thread the disc formed after a red giant star is disrupted, which can then be advected inwards by the inflow gas to the BH. In principle, the inner disc with corona depleted by the episodic accretion event can be calculated when all the detailed physics of the interaction between them are properly considered, however, this is beyond the scope of this paper.

The timescale of the episodic accretion event sweeping up the inner thin disc is estimated in Section \ref{sec:impact_acc}. In our model, when the X-ray emission reaches the dip of its light curve, it corresponds to the accretion front of the episodic event approaching the ISCO exactly, i.e., the whole inner thin disc with corona is just pushed into the BH by the gas in the episodic accretion event. At this moment, the structure of such an accretion disc without magnetic field is calculated with a supercritical accretion disc model \citep{2022ApJ...936..141C}, when the mass accretion rate $\dot{m}_{\rm epi}$ at $r_{\rm epi}$ and the value of viscosity parameter $\alpha$ are specified. With the derived disc structure, we tentatively calculate the field advection/diffusion in this disc using the method described in the previous works \citep[][]{1994MNRAS.267..235L,2011ApJ...737...94C}, when the magnetic Prandtl number $P_{\rm m}\equiv \eta/\nu$ ($\eta$ is the magnetic diffusivity and $\nu$ is the turbulent viscosity) and the strength of the external field supplied by the gas of the disrupted star at $r_{\rm epi}$ is specified.

In the presence of a large-scale magnetic field co-rotating with the gas in the disc, the outflows would be accelerated by the radiation force together with magnetic field \citep[][]{2014ApJ...783...51C}. Here we estimate the magnetic torque exerted on the unit area of the disc by the outflows \citep[][]{2013ApJ...765..149C},
\begin{equation}
T_{\rm m}={\frac {B_zB_{\phi}^{\rm s}}{2\pi}}R,\label{t_m_1}
\end{equation}
where $B_z$ and $B_{\phi}^{\rm s}$ are vertical and azimuthal components of the large-scale magnetic field at the disc surface, respectively. In the case of dense slow outflow, $B_{\phi}^{\rm s}\sim {B_z}$ \citep*[see the detailed discussion in][]{1999ApJ...512..100L,2014ApJ...783...51C}.

Part of the angular momentum and the gravitational energy released in the disc are tapped into the outflows, which increases its radial velocity, and reduces the disc surface density and the radiation. Using Equation (\ref{t_m_1}), we obtain the radial velocity of the disc with magnetic outflows as
\begin{equation}
    v_R^\prime=(1+f_{\rm m})v_{R,\rm vis}, \label{v_r}
\end{equation}
where $v_{R,\rm vis}$ is the radial velocity of a viscously driven disc (hereafter, we use the superscript ``~$\prime$~" to denote the quantities of the disc with magnetically driven outflows). The factor
\begin{equation}
f_{\rm m}={\frac {v_{R,\rm m}}{v_{R,\rm vis}}}, ~~~~{\rm and}~~~~ v_{R,\rm m}={\frac {2T_{\rm m}}{\Sigma R\Omega}},
\label{f_m}
\end{equation}
where $\Sigma$ and $\Omega$ are the surface density and angular velocity of the disc \citep[][]{2013ApJ...765..149C,2016ApJ...833...30C}. The radiation flux $f_{\rm rad}^\prime$ from the unit area of the disc with magnetic outflows is
\begin{equation}
    f_{\rm rad}^\prime=(1+f_{\rm m})^{-1}f_{\rm rad},\label{f_rad}
\end{equation}
in the presence of the magnetically driven outflows \citep[][]{2016ApJ...817...71C}, where $f_{\rm rad}$ is the flux without magnetic field.

The advection/diffusion of large-scale (poloidal) magnetic fields in the disc with magnetic outflows is described by
\begin{equation}
{\frac {\partial}{\partial t}}[R\psi(R,0)]=-v_{R}^\prime{\frac
{\partial}{\partial R}}[R\psi(R,0)]-{\frac {4\pi\eta}c}R{\frac {J_\phi^{\rm s}}{2H^\prime}}, \label{indu_1}
\end{equation}
where $J_\phi^{\rm s}$ is the vertical integrated azimuthal current density, $\psi(R,z)$ is the azimuthal component of the magnetic potential, and $\eta$ is the magnetic diffusivity \citep*[][]{1994MNRAS.267..235L}. It can be re-written as
\begin{equation}
{\frac {\partial}{\partial t}}[R\psi(R,0)]=-v_{R}^\prime{\frac
{\partial}{\partial R}}[R\psi(R,0)]-\alpha c_{\rm s}^\prime P_{\rm m}R B_R^{\rm s}, \label{indu_2}
\end{equation}
where the magnetic Prandtl number $P_{\rm m}\equiv \eta/\nu$, $B_R^{\rm s}=2\pi J_\phi^{\rm s}/c$ and the viscosity $\nu=\alpha c_{\rm s}^\prime H^\prime$ are adopted.

The configuration/strength of the large-scale magnetic fields is available with the derived potential $\psi$:
\begin{equation}
B_R(R,z)=-{\frac {\partial }{\partial z}}\psi(R,z),\label{b_r}
\end{equation}
and
\begin{equation}
B_z(R,z)={\frac {1}{R}}{\frac {\partial}{\partial R}}[R\psi(R,z)].
\label{b_z}
\end{equation}

As a substantial fraction of the angular momentum of the gas in the disc is removed by the magnetic outflows, the disc structure (both the radial velocity and the disc thickness) is significantly different from that of the disc without a magnetic field. Direct self-consistent calculations of such a disc coupled with outflows were tentatively carried out by \citet{2019ApJ...872..149L}, which are still limited to the thin disc case with a radially constant relative disc thickness $H/R$ as an input parameter.

In the absence of a magnetic field, the structure of a disc accreting at a high rate without a magnetic field is calculated by  \citet{2022ApJ...936..141C}. The radial velocity of such a disc is related to that of the disc with magnetically driven outflows as
\begin{equation}
     v_R^\prime=(1+f_{\rm m})v_{R,\rm vis}\simeq-{\frac 3 2}(1+f_{\rm m})\alpha c_{\rm s}^\prime \left({\frac {H^\prime}R}\right),  \label{v_r_2}
\end{equation}
where
\begin{equation}
H^\prime=(1+f_{\rm m})^{-1}H,~~~~{\rm and}~~~~c_{\rm s}^\prime=(1+f_{\rm m})^{-1}c_{\rm s}, \label{h_c_s}
\end{equation}
because the disc thickness is proportional to the radiation flux $f_{\rm rad}$ for a radiation pressure dominant disc, and $c^\prime_{\rm s}\simeq H^\prime \Omega_{\rm K}$. Here Equation (\ref{f_rad}) is adopted. Substituting Equations (\ref{v_r_2}) and (\ref{h_c_s}) into Equation (\ref{indu_2}), we arrive at
\begin{equation}
(1+f_{\rm m}){\frac {\partial}{\partial t}}[R\psi(R,0)]=-v_{R}{\frac
{\partial}{\partial R}}[R\psi(R,0)]-\alpha c_{\rm s} P_{\rm m}R B_R^{\rm s}. \label{indu_3b}
\end{equation}
For steady case, $\partial R\psi/\partial t=0$, we note this equation has the same mathematical form as Equation (\ref{indu_2}) describing the field advection in the disc with magnetic outflows. It implies that the field advection/diffusion in the disc with magnetic outflows can be fairly well described by solving the induction equation based on the disc structure calculated with the supercritical disc model, i.e., the complexity of the calculations of the disc-outflow couple can be avoided if we focus on the steady case. We briefly summarize the calculations of the field advection in a disc in the following paragraphs \citep*[see][for the details]{2011ApJ...737...94C}.

In the case of an external field dragged inwards by an accretion disc, the magnetic field potential $\psi(R,z)=\psi_{\rm d}(R,z)+\psi_{\infty}(R,z)$, where $\psi_{\rm d}(R,z)$ is contributed by the currents in the accretion flow, and $\psi_{\infty}(R)=B_{\rm ext}R/2$ is the external imposed homogeneous vertical field, which can be regarded as being contributed by the currents at infinity \citep[see][for the details]{1994MNRAS.267..235L,2011ApJ...737...94C}. The potential $\psi_{\rm d}$ is related to $J_{\phi}^{\rm s}(R)$ with
\begin{displaymath}
\psi_{\rm d}(R,z)={\frac {1}{c}}\int\limits_{R_{\rm ISCO}}^{R_{\rm epi}}R_{\rm d}{\rm
d}R{\rm d}\int\limits_{0}^{2\pi}\cos\phi{\rm d}\phi
\end{displaymath}
\begin{displaymath} 
\times \int\limits_{-H}^{H}
 {\frac {J_\phi(R_{\rm d},z_{\rm h})}{[{R_{\rm d}^2+R^2+(z-z_{\rm
h})^2-2R{R_{\rm d}\cos{\phi}]^{1/2}}}}}{\rm d}z_{\rm h}.
\end{displaymath}
\begin{displaymath}  
={\frac {1}{2H^\prime c}}\int\limits_{R_{\rm ISCO}}^{R_{\rm epi}}J_\phi^{\rm s}(R_{\rm d})R_{\rm d}{\rm d}R_{\rm d}\int\limits_{0}^{2\pi}\cos\phi{\rm d}\phi
\end{displaymath}
\begin{equation}
\times\int\limits_{-H}^{H}{\frac {1}{[{R_{\rm d}}^2+R^2+(z-z_{\rm
h})^2-2R{R_{\rm d}}\cos{\phi}]^{1/2}}}{\rm d}z_{\rm h}.
\label{psi}
\end{equation}
Together with this equation, the induction equation (\ref{indu_3b}) is closed. 

It is difficult to model the time evolution of the field advection in the gas of episodic accretion event moving into the inner disc with corona. In fact, only the timescale of such interaction has been estimated in a rather simplified way (see Section \ref{sec:impact_acc}). When the gas driven by the episodic accretion event reaches the ISCO, it is plausible to assume that the field of the accretion disc with magnetic outflows can be described by the induction equation (\ref{indu_2}) by letting  $\partial/\partial t=0$, which becomes
\begin{equation}
-{\frac {\partial }{\partial R}}[R\psi_{\rm d}(R,0)]-{\frac
{2\pi}{c}}{\frac {\alpha c_{\rm s}R}{v_R}}{P}_{\rm
m}J_{\phi}^{\rm S}(R)=B_{\rm ext}{R},\label{indu_3}
\end{equation}
where the structure of an accretion disc without a magnetic field is adopted as aforementioned discussion. The configuration and the strength of an external coherent magnetic field $B_{\rm ext}$ advected inwards by a disc are available by solving an integral differential equation (\ref{indu_3}) together with Equations (\ref{b_r}), (\ref{b_z}), and (\ref{psi}), based on the disc structure without magnetic field when the specified values of the disc parameters and the external field strength $B_{\rm ext}$ are specified \citep[see][for the details of the calculations]{2011ApJ...737...94C}. The magnetic Prandtl $P_{\rm m}\sim 1$ is expected for isotropic turbulence \citep[][]{1979cmft.book.....P},
which is consistent with the results of the numerical simulations, $P_{\rm m}=0.2-1$ \citep[][]{2003A&A...411..321Y,2009A&A...507...19F,2009ApJ...697.1901G,2009A&A...504..309L}. In this work, we adopt $P_{\rm m}=0.5$ in all of the calculations. With the derived field strength, the magnetic torque exerted by the outflows, and then the fraction of the power removed by the outflows are calculated with Equations (\ref{t_m_1}) and (\ref{f_m}). The radiation flux of the disc with magnetic outflows is available with Equation (\ref{f_rad}). Integrating over the disc surface, the spectrum of the disc is calculated, in which an empirical color correction for the blackbody emission from the disc has been employed \citep[][]{2001ApJ...559..680H,2002ApJ...572...79C}.

There is evidence of a rapidly spinning BH in 1ES~1927+654 \citep[][]{2022ApJ...934...35M}. The disc model with magnetically driven outflows employed in this work is in the Newtonian frame, which in principle is unable to model the spectrum of a disc surrounding a spinning BH. An accretion disc-outflow model in the general relativistic frame is beyond the scope of this work. Here we simply shift the inner edge of the disc to a radius smaller than $6GM_{\rm bh}/c^2$ in our model calculations to mimic the structure and then the spectrum of the disc surrounding a spinning BH. The inner radius of the disc, $R_{\rm in}=3R_{\rm g}$, is adopted in all of the model calculations.

\subsection{Model fitting of the X-ray spectra}\label{sec:x_ray_spect_fit}

As described in Section \ref{sec:acc_outflow}, the spectral calculations for the disc with magnetic outflows can be carried out provided the values of parameters, $\alpha$, $\dot{m}$, $\dot{m}_{\rm epi}$, $r_{\rm epi}$, and $B_{\rm ext}$ are given. The accretion rate of the inner disc is estimated as $\dot{m}=0.15$ with the X-ray observations in the pre-outburst phase, while the accretion rate $\dot{m}_{\rm epi}=6$ of the episodic accretion event in the outer region is estimated with the ratio of the UV/optical outburst peak flux to that in the pre-outburst phase (see Section \ref{sec:impact_acc} for the details). For a given value of $\alpha$, the time delay $\delta\tau\sim 150$~days of the X-ray dip to the UV/optical outburst sets a constraint on the location of the episodic accretion event (see Equation \ref{delta_tau} and Figure \ref{fig:fig1}), which gives $r_{\rm eps}\simeq 350$, and 480, for $\alpha=0.03$, and 0.1, respectively. The external field strength $B_{\rm ext}$ is a parameter to be derived with the spectral fitting of the soft X-ray emission during the X-ray dip phase.

{Here we will testify the disc-outflow model for the X-ray dip against the observations. In order to implement our disc-outflow model into the framework of XSPEC \citep{1996ASPC..101...17A}, we construct a table model \texttt{ADMDO} (accretion disc with magnetically driven outflows) in a fits file for which each individual spectrum is calculated for a given value of the external field strength $B_{\rm ext}$. We consider two cases of the table model: $\alpha=0.03$ (M1) and $0.1$ (M2). Note that, for both two table models, the black hole mass and the redshift, are specified in the model calculations. Thus, the normalization of M1 and M2 in XSPEC should be frozen to unity, and the only free parameter of \texttt{ADMO} is $B_{\rm ext}$. Furthermore, we also consider the galactic absorption (\texttt{tbabs} in XSPEC, allowing the hydrogen density to vary as a fitting parameter). So, our overall fitting model is \texttt{tbabs*ADMDO} in XSPEC notation.}

The X-ray dip roughly occurs in July-August 2018, according to the high-cadence X-ray monitoring \citep{2020ApJ...898L...1R}. We first produce the clean event data by using the tool \texttt{nicerl2} in the software package NICERDAS (version 10) with the calibration version of xti20221001. We then extract the spectra with \texttt{nicerl3-spect} and choose the SCORPEON background file \footnote{\url{https://heasarc.gsfc.nasa.gov/docs/nicer/analysis_threads/scorpeon-overview/}}. The spectra are optimally rebinned with minimum counts of 10 per bin \citep{2016A&A...587A.151K}. It turns out that the majority of the observations are dominated by the background in the sense that the background-subtracted counts rate is lower than 50\% of the background counts rate, except for two epochs: MJD 58323 (ObsID: 1200190151) and MJD 58325 (ObsID: 1200190153). The spectra of these two epochs are then combined for better statistics using the tool \texttt{addspec}\footnote{\url{https://heasarc.gsfc.nasa.gov/lheasoft/ftools/fhelp/addspec.html}}.

\begin{figure}
	\centering
	\includegraphics[width=1.0\columnwidth]{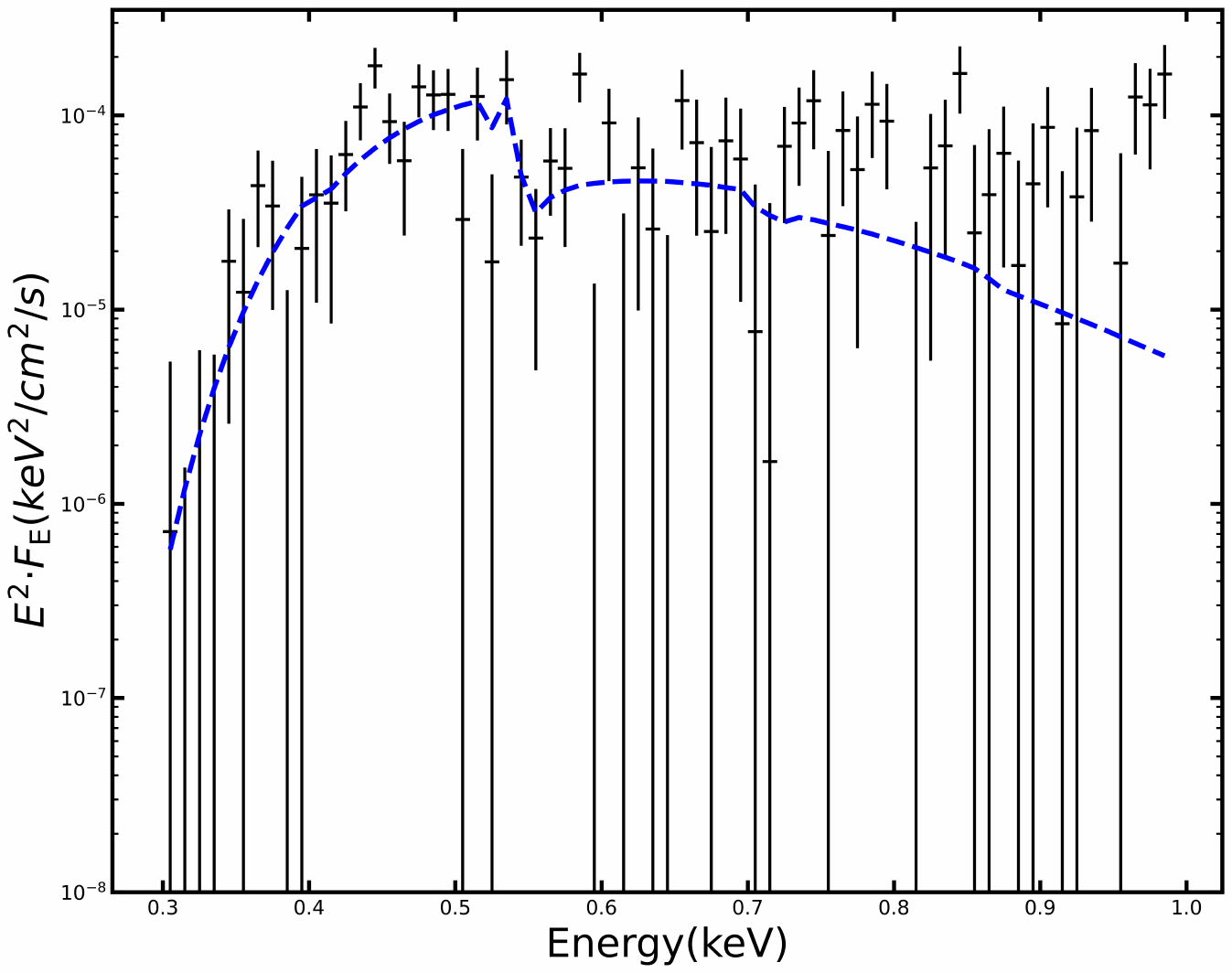}
	\caption{Unfolded energy spectrum by NICER with the best-fitting of \texttt{tbabs*ADMO} model for $\alpha=0.1$.}
	\label{fig:a0p1}
\end{figure}

{The combined energy spectrum is then fitted by the fitting model described above, which resulted in the same $N_{\rm h} = 3.1\times 10^{21} \rm cm^{-2}$ and the C-Statistic of 110.19 (69 degrees of freedom) for both M1 and M2. In Figure \ref{fig:a0p1}, we plot the best-fitting energy spectrum for M2, as an illustration of the spectral fitting. We note that, the best-fitting values of $B_{\rm ext}=2.45\times 10^3~{\rm Gauss}$ ($\beta_{\rm ext}=316$, $\beta_{\rm ext}=8\pi p/B_{\rm ext}^2$ at $r_{\rm epi}$) for M1 ($\alpha=0.03$), while $2.19\times 10^3~{\rm Gauss}$ ($\beta_{\rm ext}=85.5$) for M2 ($\alpha=0.1$). In both cases, the estimated luminosity of the accretion disc in 0.3-2 keV is about $1.0 \times 10^{42} \rm erg/s$, roughly agreeing with the one in \cite{2021ApJS..255....7R} which implements the \texttt{bbody} in XSPEC to account for the thermal soft X-ray emission.}

\begin{figure}
	\centering
	\includegraphics[width=1.0\columnwidth]{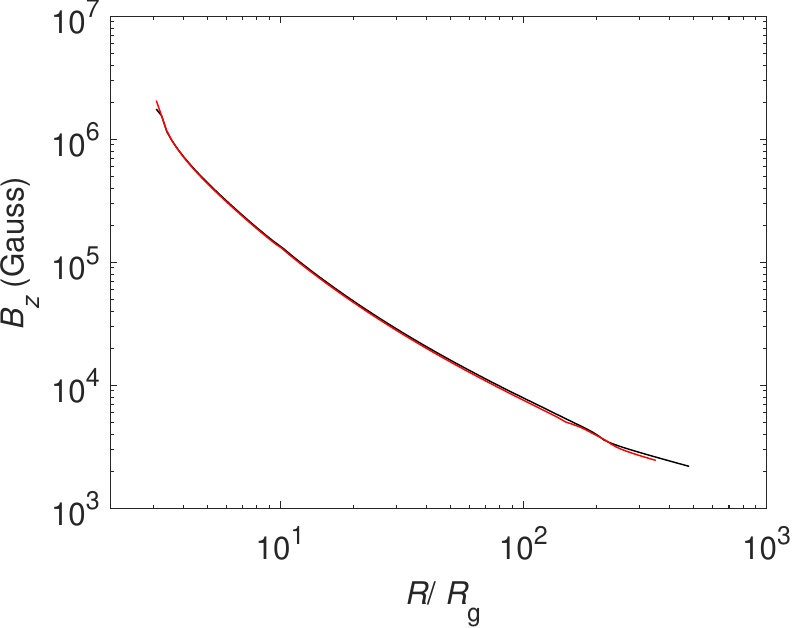}
	\caption{The field strengths vary with a radius of the disc for the best-fitted model calculations. The black line is the result for $\alpha=0.1$, while the red one is for $\alpha=0.03$. }
	\label{fig:fig3}
\end{figure}

In Figure \ref{fig:fig3}, we plot the field strengths varying with the disc radius corresponding to the best spectral fitting of the spectrum during the X-ray dip period. The external field strength, $B_{\rm ext}=2.45\times 10^3~{\rm Gauss}$ and $2.19\times 10^3~{\rm Gauss}$, for $\alpha=0.03$ and $0.1$, respectively.

\subsection{Disappearance and reappearance of the disc with corona}\label{sec:reapp_disc_cor}

It was suggested that the magnetic field can be dragged inwards by the hot gas (corona) above the disc \citep[][]{2009ApJ...701..885L,2009ApJ...707..428B,2012MNRAS.424.2097G,2013MNRAS.430..822G}, however, the field strength is limited by the gas pressure of the corona \citep[][]{2018MNRAS.473.4268C}. If the magnetic pressure surpasses the gas pressure of the corona, the hot gas in the corona may be driven by the magnetic field lines co-rotating with the disc. Thus, the corona may be destructed by the field when its strength is larger than a critical value $B_{\rm crit}$, which can be estimated by letting
\begin{equation}
  {\frac {B_{\rm crit}^2}{8\pi}}\sim p_{\rm cor}. \label{b_crit}
\end{equation}
Assuming vertical hydrostatic equilibrium in the corona, the thickness of the corona
\begin{equation}
    H_{\rm cor}=\left({\frac {p_{\rm cor}}{\rho_{\rm cor}}}\right)^{1/2}\Omega_{\rm K}^{-1}, \label{h_cor}
\end{equation}
and the optical depth $\tau_{\rm cor}=H_{\rm cor}\kappa_{\rm T}$. Substituting Equation (\ref{h_cor}) into (\ref{b_crit}), we obtain
\begin{equation}
    B_{\rm crit}\sim 5.03\times 10^8 \tau_{\rm cor}^{1/2}\tilde{H}_{\rm cor}^{-1/2}m^{-1/2}r^{-1}~~{\rm Gauss},\label{b_crit2}
\end{equation}
where $\tilde{H}_{\rm cor}=H_{\rm cor}/R$.

\begin{figure}
	\centering
	\includegraphics[width=1.0\columnwidth]{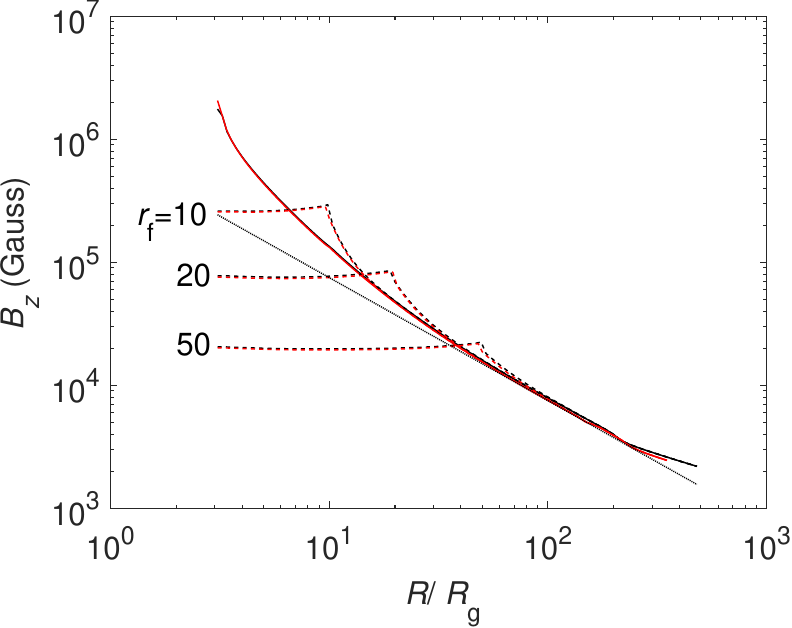}
	\caption{The field strengths varying with a radius for a slim disc connecting to the inner thin disc at $R_{\rm f}$. The solid lines are the cases of a slim disc extending to the ISCO, i.e., the same as those in Figure \ref{fig:fig3}. The black lines are the results for $\alpha=0.1$, while the red lines are for $\alpha=0.03$. The dotted line denotes the critical magnetic field defined by Equation (\ref{b_crit2}), in which typical values of the corona, $\tau_{\rm cor}=0.2$, and $\tilde{H}_{\rm cor}=0.3$, are adopted \citep[e.g.,][]{2009MNRAS.394..207C}. }
	\label{fig:fig4}
\end{figure}

As the front of the gas in the accretion event moves inwards, the outer episodic accretion disc connects to an inner thin disc with a corona at a certain radius $R_{\rm f}$, which shrinks till it approaches the ISCO (see the discussion in Section \ref{sec:impact_acc}). The field is dragged inwards by the episodic accretion disc formed after the star is disrupted. The field advection can be calculated with the induction equation (\ref{indu_3}), while the inner disc structure ($R<R_{\rm f}$) is given by the standard thin disc model, and the outer region with $R\ge R_{\rm f}$ is described by the supercritical accretion disc model.  In Figure \ref{fig:fig4}, we plot the field strengths of the inner thin disc connecting to an outer episodic accretion disc with different values of $R_{\rm f}$. It is found that the field strength is higher than the critical field strength in the inner thin disc when $r_{\rm f}\la 10$, which means the strong magnetic field is able to destruct the corona above the disc before the front of the gas in the episodic accretion event approaches the ISCO  (see panel c in Figure \ref{fig:model}). It is indeed consistent with the observations that the power-law hard X-ray component disappears before the lowest flux in the X-ray dip \citep[][]{2021ApJS..255....7R}.

The accretion rate of a TDE decays with time (most probably $\propto t^{-5/3}$) after its peak rate \citep[][]{1988Natur.333..523R}. For the outburst in 1ES~1927+654, its X-ray dip is about 150 days later than the UV/optical peak due to the TDE, which means the gas feeding rate at the outer region of the disc has declined to a rather low level. It is well known that the field advection in a thin viscous disc is inefficient due to its low radial velocity \citep[][]{1994MNRAS.267..235L}. Therefore, the field advection in the late stage after the X-ray dip observed should not be efficient as that in the early stage of the disc formed in the TDE. Accordingly, the fraction of the gravitational energy magnetically tapped into outflows decreases, and therefore the disc luminosity may increase with decreasing accretion rate. The inner field strength may decrease to a value lower than the critical value required to destruct the corona, and the accretion rate has also declined to the regime of a geometrically thin disc. The normal thin disc with corona reappears, and thus the power-law hard X-ray emission increases to the pre-outburst value or even higher \citep[][]{2021ApJS..255....7R,2022ApJ...931....5L}.

\section{Discussion}\label{sec:discuss}

1ES 1927+654 was a narrow-line Seyfert galaxy in the pre-outburst period, and its central BH was surrounded with a thin accretion disc with corona as inferred from the X-ray observations \citep[][]{2003A&A...397..557B,2013MNRAS.433..421G}. A UV/optical outburst was detected at the end of 2017, and it exhibits a TDE-like light-curve after the outburst \citep[][]{2019ApJ...883...94T}. For a normal solar-like star captured by a massive BH, its tidal disruption radius is very close to the ISCO (see Equation \ref{r_t}). Thus, the primary disc-corona in this source must be destructed instantly after such a TDE. A newly formed disc due to such a TDE should be very small in size, of which a TDE-like soft X-ray light curve is expected as those observed in normal TDEs \citep[e.g.,][]{2016MNRAS.455.2918H,2018MNRAS.474.3593K}. Unfortunately, no X-ray observation was carried out at the moment of the UV/optical outburst, whereas the disc-corona features, i.e., the thermal soft and hard X-ray components, have been detected at almost the same level of the flux as that observed in 2011 after $\sim 100$ days after the UV/optical outburst, which is quite different from a normal TDE. Therefore, we suggest that an episodic  disc accreting at a high rate formed after the TDE in the outer region of the disc is responsible for the outburst, where the radiation dominates in the UV/optical wavebands. It leads to a strong UV/optical outburst. If this is the case, the disrupted star must be large in size or/and in mass (see Figure \ref{fig:fig2}). The candidate for a star with a large radius can be either a red giant star or a proto-star. The duration of a solar mass red giant star is $\sim 1~{\rm Gyr}$, nearly one-tenth of its main sequence duration \citep[][]{1992A&AS...96..269S}. Thus, the chance of a red giant star disrupted by a BH should be comparable with (if not higher than) that of a main sequence star, since the probability of a star moving into the region within several hundred gravitational radii from the BH is much higher than that near the ISCO. We note that a star in the accretion disc evolves more rapidly than an isolate star \citep[e.g.,][]{2021ApJ...910...94C,2021ApJ...911L..14W,2021ApJ...916...48D,2023MNRAS.520.4502W}, which may increase the chance of the star disrupted by the BH. Another candidate is a proto-star, of which the size is also significantly larger than that of its main sequence counterpart, however, the duration of a proto-star is much shorter than that of a red giant star \citep[][]{1961PASJ...13..450H,1996A&A...307..829B}. It means the chance of a proto-star disrupted in this source is extremely low.

The rising timescale of the UV/optical outburst in 1ES 1927+654 is $\sim 50$ days \citep[e.g.,][]{2019ApJ...883...94T}. The characteristic fall-back time for the debris of a TDE can be estimated by
\begin{equation}
t_{\rm fb}=0.36\left({\frac {R_{\rm t}}{R_{\rm p}}}\right)^{-3n/2}\left ({\frac {M_{\rm bh}}{10^7 M_\odot}}\right )^{1/2}\left({\frac {M_*}{M_\odot}}\right)^{-1}\left({\frac {R_*}{R_\odot}}\right)^{3/2} {\rm yr},
\label{t_fb}
\end{equation}
where $R_{\rm p}$ is the orbit pericentric radius of the star, and the parameter $n$ describes the spread in orbit binding energy of the disrupted debris \citep[][]{2013MNRAS.435.1809S,2016MNRAS.461..371K}. For 1ES 1927+654, the fall-back time $t_{\rm fb}=50$~days for a solar mass red giant star with 5 solar radii if  $R_{\rm p}=0.6R_{\rm t}$, as $n=2$ is suggested in \citet{2013MNRAS.435.1809S}. {For 1ES 1927+654, the fall-back time $t_{\rm fb}=50$~days would be expected for a solar mass red giant star with 5 solar radii, if $R_{\rm p}=0.6R_{\rm t}$ and $n=2$, as is suggested in \citet{2013MNRAS.435.1809S}.}    

The gas front of such an episodic accretion event moves inwards, which gradually depletes the inner accretion disc-corona. The energy transport interior a red giant is mainly via convection, which also helps generate a strong magnetic field through dynamo processes \citep[][]{1971ApJ...164..491P,1974ApJ...193..419L}.  As the magnetic field of the red giant star is strong \citep[][]{2007ApJ...671..802B,2015Sci...350..423F}, the field may be dragged by the gas in the disc formed after the TDE, which can be substantially amplified and is co-rotating with the disc to accelerate the gas into the outflows. The radiation of such a disc with magnetic outflows could be faint if most of the gravitation energy of the viscously dissipated in the disc is tapped into the outflows \citep[][]{2013ApJ...765..149C,2016ApJ...833...30C}. 

The formation of the corona may be due to the evaporation of the cold disc \citep[][]{2000A&A...361..175M}, which may probably be heated by the re-connection of the field lines emerged from the cold disc \citep[][]{1979ApJ...229..318G,1991ApJ...380L..51H}. {The corona is thought to be formed due to the evaporation of the cold disc \citep[][]{2000A&A...361..175M}, which be probably heated by the re-connection of the field lines emerged from the cold disc \citep[][]{1979ApJ...229..318G,1991ApJ...380L..51H}.} These processes also take place in the accretion disc with magnetic outflows, however, the strong magnetic field may accelerate the hot gas into coronal outflows as those observed in some AGNs \citep[e.g.,][]{2014ApJ...783..106L}. As the field is not locally confined by the disc, instead it is maintained by the currents mostly in the outer episodic disc, so the field strength in the thin disc-corona region with $r<r_{\rm f}$ increases substantially with the accretion front moving inwards (see Figure \ref{fig:fig4}). The hot gas in the corona may be magnetically driven into outflows when the front of the gas driven by the episodic accretion event approaches to the region $\sim 10R_{\rm g}$. If the evaporation rate of the gas into the corona is less than the mass loss rate in the outflows, the corona and then its hard X-ray emission may be significantly suppressed, even before the whole inner thin disc has been swept by the episodic accretion event. 

The observed X-ray spectral evolution can in principle set further constraints on our model, if the interaction of the episodic accretion event with the inner thin disc corona is delicately modeled, most probably with numerical simulations, which is beyond the scope of this paper. However, the observed X-ray flux reaches the lowest level when the accretion front approaches the ISCO, which provides a useful constraint on our model. 
We found that the spectrum can be well fitted by our model if the external magnetic field strength is $\sim 2000$~Gauss, which is consistent with the scenario of a red giant star disrupted by the BH, because the typical field strengths of red giant stars are several thousand Gauss in their envelopes, or $\sim 10^5$~Gauss in the cores of the red giant stars \citep[][]{2007ApJ...671..802B,2015Sci...350..423F}. Although our model of an accretion disc with magnetic outflows is in the Newtonian frame, we tentatively extend it for a spinning BH by shifting the ISCO to a small radius, which is rather simplified. Nevertheless, the present X-ray observations are still insufficient to constrain the detailed temperature distribution of the disc near the ISCO, which in fact is only able to provide a rough estimate of the characteristic temperature of the disc. Thus, we believe present spectral fitting on the X-ray observations may still set fairly good constraints on the model (see Figure \ref{fig:a0p1}).

The origin of the soft X-ray emission is still under debate, and one possibility is the soft X-ray excess ubiquitously observed in normal AGNs \citep[e.g.,][and the references therein]{2004MNRAS.349L...7G,2012MNRAS.420.1848D}. The temperature of the soft X-ray excess in AGNs does not vary with the source luminosity \citep[][]{2004MNRAS.349L...7G,2012MNRAS.420.1848D}, while the blackbody temperature of this source varies from $\sim 50~{\rm eV}$ to several hundred eV, and the luminosity increases with temperature roughly as a power-law dependence \citep[][]{2021ApJS..255....7R}, which is consistent with the disc origin for the soft X-ray emission in this source, i.e., $L\propto T^4$ for a thin disc \citep[][]{1973A&A....24..337S}, or $L\propto T^2$ for a slim disc \citep[][]{2000PASJ...52..133W}. This justifies our assumption of the disc origin of the soft X-ray emission from this source.

{There is observational evidence of the outflows in this source. The optical observations show that broad Balmer lines emerged $\sim 100$ days after the outburst \citep[][]{2022ApJ...933...70L}. They suggested that the broad line region (BLR) was intrinsically absent before the optical outburst possibly due to the low column density of the clouds. The emergence of the newly formed BLR may originate in the condensed gas of the outflows driven from the disc formed after the TDE. More direct evidence of the outflows may be the X-ray observations. It was found that, a prominent, broad 1 keV feature in the X-ray spectra, can be modelled with blueshifted reflection from a single-temperature blackbody irradiating spectrum \cite[][]{2021ApJS..255....7R,2022ApJ...934...35M}. This 1 keV feature is proposed to arise from reflected emission off the base of an outflow from the disc, which is consistent with our model of an accretion disc with magnetically driven outflows.} Such observations may help constrain the model if the velocity, density, and temperature distributions of the outflows are derived with our accretion-outflow model. However, in this work, we focus on the disc structure that is impacted by the outflows, while the detailed dynamics of the outflows have not been included. The dynamics of magnetically driven outflows from a thin accretion disc with properly considering the disc-outflow couple has been explored by \citet{2022ApJ...926...11L}, which shows the terminal outflow velocity can be $\sim 0.01-0.1~{\rm c}$, which is qualitatively consistent with the X-ray observations of this source. Recent numerical simulations of the magnetically driven outflows may also help understand the observations of 1~keV features in this sources \citep[e.g.,][]{2018ApJ...867..100Y,2019ApJ...881...34Y,2022MNRAS.513.5818W,2022MNRAS.515.5594W}.

An accretion disc with strong magnetically driven outflows is a key ingredient of our model for 1ES~1927+654, and a rapidly spinning BH is preferred by the X-ray observations \citep[][]{2021ApJS..255....7R}, which implies that jets may probably be accelerated through the Blandford-Znajek mechanism or/and Blandford-Payne mechanism during the X-ray dip \citep[][]{1977MNRAS.179..433B,1982MNRAS.199..883B}. Unfortunately, no radio observation was performed during the period of the X-ray dip. The high-resolution radio observations have only been carried out before or after the outburst. Only central point source was detected by the observations before the outburst (January 1992, June 1998, and March 2014), whereas the extended emission has been detected by the observations carried out in December 2018 and March 2021 \citep[see][for the details]{2022ApJ...931....5L}. We conjecture that it may be the evidence of the jets accelerated by the magnetic field advected inwards by the gas of the episodic accretion event. Modelling the detailed radio emission with the jet formation mechanism is beyond the scope of this paper.

{In our model, a strong magnetic field of the disrupted star is required, which may not be the case for most other TDEs. If the field is weak or even without a magnetic field, strong outflows can still be launched from a super-Eddington accretion disc formed after a TDE, as shown in the numerical simulations 
\citep[][]{2018ApJ...859L..20D,2021MNRAS.507.3207C,2023MNRAS.523.4136B}. In this case, the radiation of the disc should be greater than $\sim L_{\rm Edd}$, and the outflows will be suppressed when the disc becomes faint. It implies that the X-ray dip observed in this source with substantially sub-Eddington luminosity  cannot be reproduced by our model if the field is weak or absent.}

{It is interesting to note that the X-ray emission of a TDE disc with magnetically driven outflows is still quite strong (in some case even super-Eddington) \citep[see Figure 12 in][]{2019MNRAS.483..565C}. We note that the accretion rates adopted in their simulations are in the range of $32-150~\dot{M}_{\rm Edd}$,  which are much higher than $\dot{M}=6~\dot{M}_{\rm Edd}$ in our model calculations. Their simulations show that the radiation efficiency $\eta_{\rm rad}$ of the disc can be as low as $\sim 0.001$ in strong magnetic field cases \citep[see Table 1 in][]{2019MNRAS.483..565C}. For 1ES 1927+654, the disc luminosity $\sim 2\times 10^{41} {\rm erg/s}$ is predicted if $\dot{m}=6$ and $\eta_{\rm rad}=0.001$ are adopted, which is roughly consistent with the luminosity of the X-ray dip. }

{The radiation of a TDE disc may be obscured by the outflows. One may speculate that the observed X-ray dip is caused by the X-ray photons from the disc blocked by the outflows. If this is the case, the temperature of the thermal soft X-ray emission from the disc should be unchanged when the flux decreases during the obscuration, whereas the X-ray observations show the temperature decreases with decreasing flux \citep[e.g.,][]{2021ApJS..255....7R}, which implies this scenario quite unlikely.}

\section*{Acknowledgements}

We are grateful to the referee for his/her insightful comments and suggestions. We thank Zhen Yan for the extraction of NICER spectra, Tinggui Wang, Luis Ho, Qingwen Wu, and Yanfei Jiang for helpful discussion/comments.  This work is supported by the NSFC (11833007, 12073023, 12233007, 12147103, 12322307, 12273026 and U1931203), the science research grants from the China Manned Space Project with No. CMS-CSST- 2021-A06, the fundamental research fund for Chinese central universities (Zhejiang University), the National Program on Key Research and Development Project 2021YFA0718500, the Natural Science Foundation of Hubei Province (2022CFB167), and the Fundamental Research Funds for the Central Universities (Wuhan University, 2042022rc0002), and the Xiaomi Foundation/ Xiaomi Young Talents Program.

%%%%%%%%%%%%%%%%%%%%%%%%%%%%%%%%%%%%%%%%%%%%%%%%%%
\section*{Data Availability}

The data underlying this article will be shared on reasonable request to the corresponding author.

%The inclusion of a Data Availability Statement is a requirement for articles published in MNRAS. Data %Availability Statements provide a standardised format for readers to understand the availability of data %underlying the research results described in the article. The statement may refer to original data generated %in the course of the study or to third-party data analysed in the article. The statement should describe and %provide means of access, where possible, by linking to the data or providing the required accession numbers %for the relevant databases or DOIs.

%%%%%%%%%%%%%%%%%%%% REFERENCES %%%%%%%%%%%%%%%%%%

% The best way to enter references is to use BibTeX:

\bibliographystyle{mnras}
%\bibliography{example} % if your bibtex file is called example.bib
\bibliography{tde_disk} % if your bibtex file is called example.bib

% Alternatively you could enter them by hand, like this:
% This method is tedious and prone to error if you have lots of references
%\begin{thebibliography}{99}
%\bibitem[\protect\citeauthoryear{Author}{2012}]{Author2012}
%Author A.~N., 2013, Journal of Improbable Astronomy, 1, 1
%\bibitem[\protect\citeauthoryear{Others}{2013}]{Others2013}
%Others S., 2012, Journal of Interesting Stuff, 17, 198
%\end{thebibliography}

%%%%%%%%%%%%%%%%%%%%%%%%%%%%%%%%%%%%%%%%%%%%%%%%%%

%%%%%%%%%%%%%%%%% APPENDICES %%%%%%%%%%%%%%%%%%%%%

%\appendix

%\section{Some extra material}

%If you want to present additional material which would interrupt the flow of the main paper,
%it can be placed in an Appendix which appears after the list of references.

%%%%%%%%%%%%%%%%%%%%%%%%%%%%%%%%%%%%%%%%%%%%%%%%%%

% Don't change these lines
\bsp	% typesetting comment
\label{lastpage}
\end{document}